\documentclass[11pt]{article}
\usepackage[margin=1.1in]{geometry}
\usepackage[T1]{fontenc}
\usepackage{lmodern}
\usepackage{microtype}
\usepackage{amsmath,amssymb,amsthm}
\usepackage{booktabs}
\usepackage{multirow}
\usepackage{listings}
\usepackage{xcolor}
\usepackage{tikz}
\usepackage{enumitem}
\usepackage[colorlinks=true,linkcolor=blue!50!black,citecolor=blue!50!black,urlcolor=blue!50!black]{hyperref}
\usepackage[capitalise,nameinlink]{cleveref}

\newtheorem{theorem}{Theorem}
\newtheorem{lemma}[theorem]{Lemma}
\newtheorem{corollary}[theorem]{Corollary}
\theoremstyle{definition}
\newtheorem{definition}[theorem]{Definition}
\theoremstyle{remark}
\newtheorem{remark}[theorem]{Remark}

\newcommand{\link}{\mathit{link}}
\newcommand{\cost}{\mathit{cost}}
\newcommand{\bd}{\operatorname{bd}}
\newcommand{\KRT}{\mathcal{K}}
\newcommand{\code}[1]{\texttt{#1}}

\title{Incremental Minimum Spanning Trees on a Napkin}
\author{Lucian Bicsi \and Adrian Bud\u{a}u}
\date{}

\begin{document}
\maketitle

\begin{abstract}
In the incremental minimum spanning forest problem, weighted edges arrive one at a time. After each insertion we must report the edge that leaves the minimum spanning forest, and queries ask for the bottleneck distance of two vertices: the smallest possible maximum edge cost on a path between them. Link-cut trees solve both in $O(\log n)$ amortized time. In 2021 we wrote the folklore \code{BottleneckTree}, a short competitive-programming structure that stores one parent pointer and one cost per vertex, and conjectured that it also runs in $O(\log n)$ amortized time. We show that this conjecture is false: on a family of inputs, its amortized cost per operation is $\Omega(\sqrt{n})$. We then present a structure with the same state, two arrays and a single loop, and prove that it runs in $O(\log n)$ amortized time per insertion and per query. Unlike the AM-tree of Ding, Gu and Sun, which attains the same bound, it stores no subtree sizes and maintains no balance invariant. For a connected graph with distinct costs, the structure can be viewed as an implicit link-cut tree on the Kruskal reconstruction tree: each rotation corresponds to exactly one preferred-child change, and each skip acts on a stale path-parent pointer. In experiments it is 1.3--2.1 times faster than the folklore \code{BottleneckTree} on insertions, up to 480 times faster on its adversarial input, and 6.3--16 times faster than an explicit link-cut tree.
\end{abstract}

\section{Introduction}\label{sec:intro}

Maintaining a minimum spanning forest (MSF) while edges arrive one at a time is a basic dynamic graph problem, known as \emph{incremental minimum spanning tree}. The graph has the vertices $V=\{0,\dots,n-1\}$ and real edge costs, starts without edges, and supports two operations.

\begin{description}[leftmargin=1.5em]
\item[$\code{AddEdge}(x,y,c)$:] insert the edge $\{x,y\}$ with cost $c$ and return the cost of the edge that leaves the MSF: the cost of the removed edge if the new edge enters the forest, $c$ if the new edge does not enter, and $\infty$ if $x$ and $y$ were in different components (then no edge leaves).
\item[$\code{Query}(x,y)$:] return the \emph{bottleneck distance} of $x$ and $y$, the minimum over all $x$--$y$ paths of the maximum edge cost on the path ($\infty$ if no path exists).
\end{description}

The problem is used for connectivity queries on temporal graphs~\cite{SongWXQZL24,DingGS25}: if each edge costs minus its arrival time, the edges that arrived after time $t$ connect $x$ and $y$ exactly when $\code{Query}(x,y)<-t$, which answers sliding-window connectivity~\cite{AndersonBT20} without deletions. The structure also maintains the MSF of an edge stream in $O(n)$ space, and $\code{Query}(x,y)$ is the level at which single-linkage clustering merges $x$ and $y$.

\paragraph{How simple can a logarithmic solution be?} The classical solution stores the forest in a link-cut tree~\cite{SleatorT83,SleatorT85} and takes $O(\log n)$ amortized time per operation, at the price of splay trees, path-parent pointers and reversal bits. Recent work uses a much simpler object: a rooted tree on the vertices, with one parent pointer and one cost per vertex, whose path maxima equal those of the MSF. The OEC-forest~\cite{SongWXQZL24} uses such a tree, but no bound better than $O(n)$ per operation is known for it. The AM-tree~\cite{DingGS25} attains $O(\log n)$ amortized time by also storing subtree sizes and enforcing an \emph{anti-monopoly} rule: no child may hold more than two thirds of its parent's subtree.

In 2021 we wrote the folklore \code{BottleneckTree}~\cite{BudauBicsi21} for competitive programming. It uses the same kind of tree, stores nothing else, and fits in a few lines of code. Its header states the bound ``probably $O(\log n)$ amortized'', which we conjectured when writing it. No proof has been given.

\paragraph{Our results.} We show that this conjecture is false: on an explicit family of inputs, the folklore \code{BottleneckTree} takes $\Omega(\sqrt n)$ amortized time per operation, and we explain informally why our analysis does not apply to it (\cref{sec:folklore}). We then give a structure with exactly the same state that provably runs in $O(\log n)$ amortized time. \Cref{tab:compare} summarizes the comparison.

\begin{table}[t]
\centering\small
\begin{tabular}{@{}l p{6.2cm} l@{}}
\toprule
structure & state & amortized time per operation \\
\midrule
link-cut tree~\cite{SleatorT85} & child and parent pointers, reversal bit and path maximum, for every vertex and every forest edge & $O(\log n)$ \\
AM-tree~\cite{DingGS25} & parent, cost and subtree size per vertex & $O(\log n)$ \\
folklore \code{BottleneckTree}~\cite{BudauBicsi21} & parent and cost per vertex & $\Omega(\sqrt n)$ (\cref{thm:folklore}) \\
this paper (\cref{fig:code}) & parent and cost per vertex & $O(\log n)$ (\cref{thm:main}) \\
\bottomrule
\end{tabular}
\caption{Structures for incremental MST and bottleneck queries. For the AM-tree we list its lazy variant; the strict variant also stores child pointers and bounds the worst-case query time. The lower bound for the folklore \code{BottleneckTree} holds on an explicit family of inputs.}
\label{tab:compare}
\end{table}

Like the AM-tree and the folklore \code{BottleneckTree}, our structure changes the tree only by two local moves: a \emph{rotation}, which swaps a vertex with its parent, and a \emph{skip}, which makes a vertex a child of its grandparent (\cref{sec:prelim}). Our structure differs from both mainly in the order in which it applies these moves when it brings a vertex to the root, and unlike the AM-tree it stores no sizes. Bringing a vertex to the root naively, by repeatedly rotating or skipping it, behaves like move-to-root in binary search trees~\cite{AllenM78} and takes linear time on simple inputs. We add two small repairs, in which the parent first moves one step up itself; they make the cost logarithmic without storing subtree sizes, and in our experiments, dropping either of them makes it linear again (\cref{rem:necessary}). The analysis uses the potential $\sum_v\log s(v)$ of the AM-tree.

We also show that, for a connected graph with distinct costs, the structure acts as an implicit link-cut tree on the Kruskal reconstruction tree, the tree of merges performed by Kruskal's algorithm (\cref{sec:lct}). Every rotation is exactly one preferred-child change, which a link-cut tree performs by splaying an auxiliary tree.

In experiments (\cref{sec:experiments}), our structure is 1.3--2.1 times faster on insertions than the folklore \code{BottleneckTree}, up to 480 times faster on its adversarial input, and 6.3--16 times faster than a link-cut tree. The lazy AM-tree variants are mostly faster than ours, by up to 2 times, when edges arrive in random order; when consecutive edges are close in the graph, ours is competitive, and on a band-shaped graph inserted in this order it is 1.1--1.6 times faster than both.

\paragraph{Outline.} \Cref{sec:prelim,sec:algorithm,sec:analysis} give the moves, the algorithm and its analysis, \cref{sec:lct} the link-cut-tree view, \cref{sec:folklore} the lower bound, and \cref{sec:experiments} the experiments.

\section{Preliminaries}\label{sec:prelim}

We assume $x\neq y$ in both operations; a self-loop never enters the MSF. Because the structures we study only compare and move costs, we can break ties by replacing the cost $c_j$ of the $j$-th inserted edge with the pair $(c_j,j)$, compared lexicographically. Costs are then distinct, and both operations can return the edge itself rather than only its cost. Only \cref{sec:lct} needs distinct costs.

We write $\bd_G(u,w)$ for the bottleneck distance of $u$ and $w$ in $G$. It equals the maximum cost on the $u$--$w$ path of any MSF of $G$.

\paragraph{Representation.}
The structure maintains a rooted tree $T$ on $V$. Each non-root vertex $v$ stores its parent $\link[v]$ and the cost $\cost[v]$ of the edge $\{v,\link[v]\}$. Initially, every vertex $v\neq0$ is a child of $0$ with the placeholder cost $\infty$, and the root has the sentinel cost $\infty^{+}$, which exceeds every other cost. This remains true: the operations below exchange costs only between a vertex and its parent, so the cost of the root moves with the root, and the cost update in $\code{AddEdge}$ acts on a non-root vertex. We assume that all inserted costs are finite. The edges of $T$ need not be edges of $G$. The structure maintains the following invariant, as do the OEC-forest and the AM-tree.

\begin{description}[leftmargin=1.5em]
\item[Invariant (I).] For all $u,w\in V$, the maximum cost on the $u$--$w$ path of $T$ equals $\bd_G(u,w)$.
\end{description}

\begin{figure}[!hb]
\centering
\begin{tikzpicture}[v/.style={circle,draw,minimum size=6.5mm,inner sep=0pt}, hx/.style={v,fill=blue!8,draw=blue!50!black,thick}, lab/.style={font=\scriptsize}, st/.style={draw=black!55}, >=stealth]
\newcommand{\subb}[2]{\draw[st] (#1.south) -- ++(-0.22,-0.48) -- ++(0.44,0) -- cycle; \path (#1.south) ++(0,-0.33) node[font=\tiny]{$#2$};}
\newcommand{\subr}[2]{\draw (#1) -- ++(0.55,-0.45) coordinate (#1t); \draw[st] (#1t) -- ++(-0.22,-0.48) -- ++(0.44,0) -- cycle; \path (#1t) ++(0,-0.33) node[font=\tiny]{$#2$};}
\begin{scope}
\node[v] (g) at (0,2.5) {$g$}; \node[v] (p) at (0,1.25) {$p$}; \node[hx] (x) at (0,0) {$x$};
\draw (g)--node[lab,left]{$c_p$}(p); \draw (p)--node[lab,left]{$c_x$}(x);
\subr{g}{C} \subr{p}{B} \subb{x}{A}
\draw[->] (0.95,1.25)--(1.85,1.25);
\node[v] (g2) at (2.8,2.5) {$g$}; \node[hx] (x2) at (2.8,1.25) {$x$}; \node[v] (p2) at (2.8,0) {$p$};
\draw (g2)--node[lab,left]{$c_p$}(x2); \draw (x2)--node[lab,left]{$c_x$}(p2);
\subr{g2}{C} \subr{x2}{A} \subb{p2}{B}
\node[align=center] at (1.4,-1.35) {rotation of $x$ ($c_x\le c_p$):\\ $x$ swaps with its parent $p$};
\end{scope}
\begin{scope}[xshift=6.6cm]
\node[v] (g) at (0,2.5) {$g$}; \node[v] (p) at (0,1.25) {$p$}; \node[hx] (x) at (0,0) {$x$};
\draw (g)--node[lab,left]{$c_p$}(p); \draw (p)--node[lab,left]{$c_x$}(x);
\subr{g}{C} \subr{p}{B} \subb{x}{A}
\draw[->] (0.95,1.25)--(1.85,1.25);
\node[v] (g2) at (3.1,2.5) {$g$}; \node[v] (p2) at (2.3,1.1) {$p$}; \node[hx] (x2) at (3.9,1.1) {$x$};
\draw (g2)--node[lab,left]{$c_p$}(p2); \draw (g2)--node[lab,right]{$c_x$}(x2);
\subb{g2}{C} \subb{p2}{B} \subb{x2}{A}
\node[align=center] at (1.6,-1.35) {skip of $x$ ($c_p\le c_x$):\\ $x$ becomes a child of its grandparent $g$};
\end{scope}
\end{tikzpicture}
\caption{The two moves. Each edge is labeled with the cost stored at its lower endpoint. The triangles $A$, $B$ and $C$ stand for the other children of $x$, $p$ and $g$ with their subtrees; they keep their parents, and no other vertex changes its parent.}
\label{fig:moves}
\end{figure}
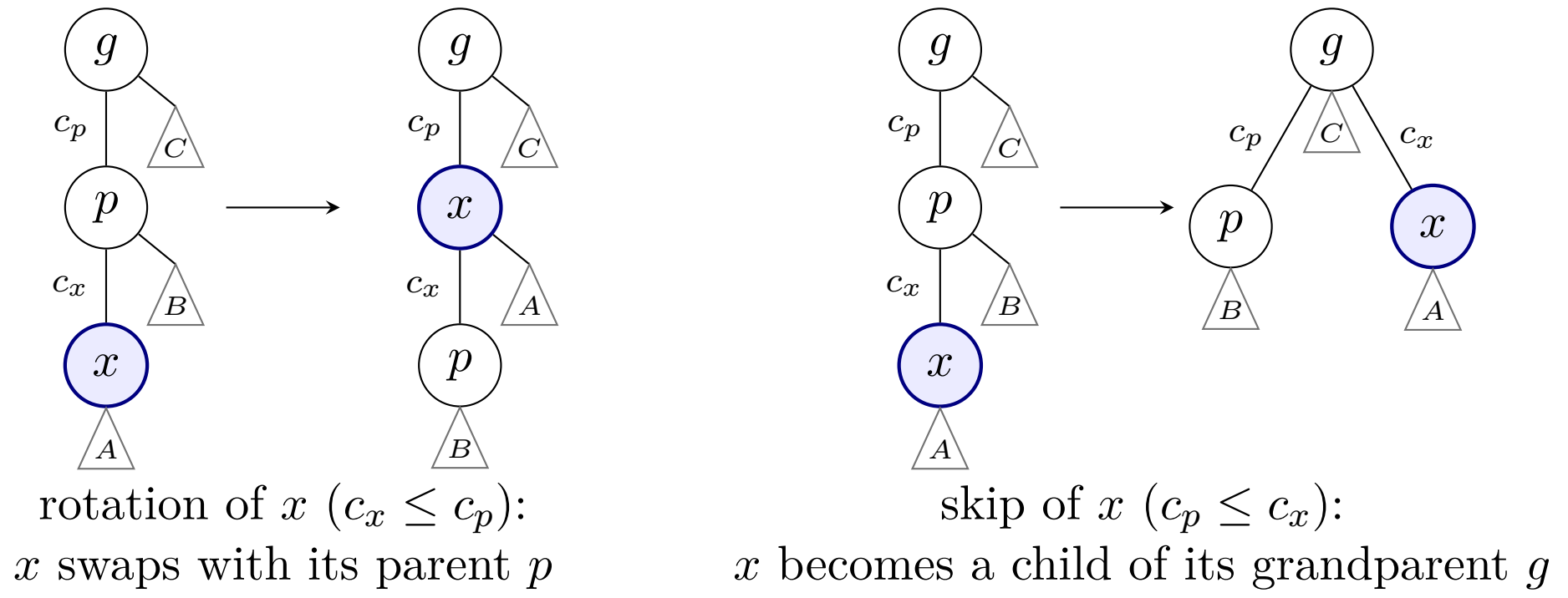

Let $x$ be a non-root vertex with parent $p$, and let $c_x=\cost[x]$ and $c_p=\cost[p]$, where $c_p=\infty^{+}$ if $p$ is the root. The edge of $x$ is an \emph{up edge} if $c_x\le c_p$ and a \emph{down edge} otherwise. We use two moves (\cref{fig:moves}), which correspond to the rotate and shortcut cases of the \emph{Promote} operation of the AM-tree~\cite{DingGS25}.

\begin{description}[leftmargin=1.5em]
\item[Rotation of $x$] (requires $c_x\le c_p$). Vertex $x$ takes the place of $p$: it becomes a child of the parent $g$ of $p$ with cost $c_p$ (or the root, if $p$ was the root), and $p$ becomes a child of $x$ with cost $c_x$. No other vertex changes its parent.
\item[Skip of $x$] (requires $c_p\le c_x$; then $p$ is not the root). Vertex $x$ becomes a child of $g$ and keeps the cost $c_x$.
\end{description}

\begin{lemma}[cf.~\cite{SongWXQZL24,DingGS25}]\label{lem:moves}
Rotations and skips preserve Invariant~(I).
\end{lemma}
\begin{proof}
A tree path that uses a modified edge enters and leaves the set $\{x,p,g\}$, so it suffices to compare the path maxima between pairs of these vertices. For a rotation, the pairs $\{x,p\}$, $\{p,g\}$, $\{x,g\}$ have maxima $c_x$, $c_p$, $\max(c_x,c_p)=c_p$ before and $c_x$, $\max(c_x,c_p)=c_p$, $c_p$ after; if $p$ is the root, only the pair $\{x,p\}$ exists and keeps the maximum $c_x$. For a skip, the maxima are $c_x$, $c_p$, $\max(c_x,c_p)=c_x$ before and $\max(c_x,c_p)=c_x$, $c_p$, $c_x$ after.
\end{proof}

Every edge is an up edge or a down edge, so a non-root vertex can always move one level up: by a rotation if its edge is up, and otherwise by a skip, since a child of the root always has an up edge.

\section{The algorithm}\label{sec:algorithm}

Procedure $\code{access}(x)$ moves $x$ to the root of $T$. Let $p=\link[x]$ and $g=\link[p]$. Each iteration of its loop performs one of four cases.

\begin{description}[leftmargin=1.5em]
\item[(A)] The edge of $x$ is an up edge: rotate $x$.
\item[(B)] The edges of $x$ and $p$ are down edges: skip $p$, then skip $x$.
\item[(C)] The edge of $x$ is a down edge, the edge of $p$ is an up edge, and $\cost[g]\le\cost[x]$: rotate $p$, then skip $x$.
\item[(D)] The edge of $x$ is a down edge, the edge of $p$ is an up edge, and $\cost[g]>\cost[x]$: skip $x$.
\end{description}

\noindent Thus, at a down edge, $p$ first takes one step of its own, provided the step does not raise $\cost[p]$ above $\cost[x]$ (in case (D), $p$ does not move), and then $x$ skips $p$. \Cref{fig:code} gives the complete implementation.

\begin{figure}[t]
\begin{lstlisting}
const int INF = 1e9;
struct BottleneckTree {
  vector<int> link, cost;   // all inserted costs must be < INF
  BottleneckTree(int n) : link(n, 0), cost(n, INF) {
    link[0] = -1; cost[0] = INF + 1;
  }
  void rot(int x) {
    int p = link[x], g = link[p];
    swap(cost[x], cost[p]);
    link[x] = g, link[p] = x;
  }
  void skip(int x) { link[x] = link[link[x]]; }
  void access(int x) {
    for (int p = link[x]; p != -1; p = link[x]) {
      int g = link[p];
      if (cost[x] <= cost[p]) rot(x);                  // (A)
      else if (cost[p] > cost[g]) skip(p), skip(x);    // (B)
      else if (cost[g] <= cost[x]) rot(p), skip(x);    // (C)
      else skip(x);                                    // (D)
    }
  }
  int Query(int x, int y) {
    access(x); access(y);
    return cost[x];
  }
  int AddEdge(int x, int y, int c) {
    access(x); access(y);
    if (cost[x] > c) swap(cost[x], c);
    return c;
  }
};
\end{lstlisting}
\caption{The complete data structure. \code{INF} represents $\infty$ and \code{INF + 1} the sentinel $\infty^{+}$. The structure keeps the name of the code it replaces (\cref{fig:folklore}).}
\label{fig:code}
\end{figure}

\begin{lemma}\label{lem:legal}
Every move performed by $\code{access}$ is a legal rotation or skip. In cases (B)--(D), $p$ is not the root, and in cases (B) and (C), $g$ is not the root.
\end{lemma}
\begin{proof}
A child of the root has an up edge, since the root has cost $\infty^{+}$. Hence in cases (B)--(D) the vertex $p$ is not the root. In case (B), $\cost[g]<\cost[p]<\infty^{+}$; in case (C), $\cost[g]\le\cost[x]<\infty^{+}$; so $g$ is not the root in either case. In case (B), $p$ has a down edge and may skip; afterwards $x$ is still a child of $p$ with $\cost[p]<\cost[x]$ and may skip. In case (C), $p$ has an up edge and may rotate; afterwards $\cost[p]$ equals the old $\cost[g]\le\cost[x]$, so $x$ may skip. In case (D), $x$ has a down edge.
\end{proof}

Each iteration decreases the depth of $x$, by one in cases (A) and (D) and by two in cases (B) and (C), so $\code{access}(x)$ terminates with $x$ at the root.

\begin{lemma}\label{lem:adjacent}
Let $x\neq y$. After $\code{access}(x)$ and $\code{access}(y)$, the root of $T$ is $y$, $x$ is a child of $y$, and $\cost[x]=\bd_G(x,y)$.
\end{lemma}
\begin{proof}
After $\code{access}(x)$, $x$ is the root and $\cost[x]=\infty^{+}$. During $\code{access}(y)$, $x$ remains the root until $y$ rotates over it: a skip of a child of $x$ would need $\cost[x]$ to be at most the child's cost, and by \cref{lem:legal} case (B) or (C) never acts on $g=x$. When $y$ rotates over $x$, $y$ becomes the root, $x$ becomes its child, and the loop ends. The last claim follows from Invariant~(I).
\end{proof}

\begin{lemma}\label{lem:addedge}
$\code{Query}(x,y)$ returns $\bd_G(x,y)$. $\code{AddEdge}(x,y,c)$ preserves Invariant~(I) and returns the cost of the edge that leaves the MSF.
\end{lemma}
\begin{proof}
The first claim is \cref{lem:adjacent}. Let $b=\cost[x]=\bd_G(x,y)$ after the two accesses, and let $e=\{x,y\}$ with cost $c$. For all $u,w$,
\[
\bd_{G+e}(u,w)=\min\bigl\{\bd_G(u,w),\ \max(\bd_G(u,x),c,\bd_G(y,w)),\ \max(\bd_G(u,y),c,\bd_G(x,w))\bigr\},
\]
so $\bd_{G+e}$ depends only on $\bd_G$ and $e$. Viewed as a weighted graph, $T$ satisfies $\bd_T=\bd_G$ by Invariant~(I), and therefore $\bd_{G+e}=\bd_{T+e}$. The only cycle of $T+e$ consists of $e$ and the edge $\{x,y\}$ of cost $b$. By the cycle property, the MSF of $T+e$ is $T$ with the cost of $\{x,y\}$ replaced by $\min(b,c)$, which is the tree after the swap. Again by the cycle property, the edge that leaves the MSF of $G+e$ is a most expensive edge of the cycle closed by $e$, whose cost is $\max(\bd_G(x,y),c)=\max(b,c)$ for any tie-breaking; this is the returned value, and it is $\infty$ if $x$ and $y$ were disconnected.
\end{proof}

\section{Analysis}\label{sec:analysis}

The cost of an access is the number $W$ of loop iterations, each of which takes constant time. For the analysis, let $s(v)$ be the number of vertices in the subtree of $v$ in $T$, and let
\[
\Phi=\sum_{v\in V}\log_2 s(v).
\]
Initially $\Phi=\log_2 n$, and $\Phi\ge 0$ always. The cost swap in $\code{AddEdge}$ changes no subtree, so $\Phi$ changes only during accesses.

\begin{lemma}\label{lem:delta}
Let $a$ be a child of $b$.
\begin{enumerate}[label=(\roman*),leftmargin=2em]
\item A rotation of $a$ changes $\Phi$ by $\log_2\frac{s(b)-s(a)}{s(a)}$. If $s(a)\ge\frac23 s(b)$, this is at most $-1$, and we call the rotation \emph{heavy}.
\item A skip of $a$ changes $\Phi$ by $\log_2\frac{s(b)-s(a)}{s(b)}\le0$. If $s(a)\ge\frac12 s(b)$, this is at most $-1$, and we call the skip \emph{heavy}.
\end{enumerate}
\end{lemma}
\begin{proof}
A rotation gives $a$ the former subtree of $b$ and leaves $b$ with the other $s(b)-s(a)$ vertices; no other subtree changes. A skip removes the subtree of $a$ from that of $b$ and changes no other subtree size.
\end{proof}

Fix an access, and let $x_0,x_1,\dots,x_d$ be the root path of the accessed vertex when the access begins, with subtree sizes $s_0<s_1<\dots<s_d=n$ at that time.

\begin{lemma}[Untouched ancestors]\label{lem:untouched}
At the start of every iteration of $\code{access}(x)$, the parent of $x$ and all its ancestors have the parent pointers and subtree sizes they had when the access began. In particular, if the parent of $x$ is not the root, the parent and grandparent of $x$ are consecutive vertices $x_i,x_{i+1}$ of the original root path.
\end{lemma}
\begin{proof}
An iteration modifies only the pointers and subtree sizes of $x$, $p$ and $g$. After the iteration, the parent of $x$ is the former $g$ in cases (A) and (D), and the former parent of $g$ in cases (B) and (C); in both cases it was not modified. The set of vertices below each ancestor of the new parent is unchanged. The claim follows by induction.
\end{proof}

By \cref{lem:untouched}, the pair $(p,g)$ in cases (B) and (C) is a pair $(x_i,x_{i+1})$ of consecutive original path vertices with their original sizes. Since $x$ moves above $g$ in these cases, different iterations use disjoint pairs.

\begin{theorem}\label{thm:main}
For every access,
\[
W+2\,\Delta\Phi\;\le\;3\log_{3/2} n+5\log_2 n\;<\;10.13\log_2 n.
\]
\end{theorem}
\begin{proof}
Each iteration is charged one unit, except that an iteration of case (D) is charged to the next iteration. The next iteration is of case (A): after the skip, the parent of $x$ is the former $g$, and $\cost[g]>\cost[x]$. Hence an iteration of case (A) carries at most two units and every other charged iteration carries one. By \cref{lem:delta}(ii), skips of $x$ do not increase $\Phi$. We bound the contribution of each case to $W+2\Delta\Phi$.

\emph{Case (A).} A heavy rotation contributes at most $2+2\cdot(-1)=0$. A light rotation increases $s(x)$ by a factor greater than $3/2$. Only case-(A) rotations change $s(x)$, so there are at most $\log_{3/2}n$ light rotations. A light rotation increases $\Phi$ by at most $\log_2\frac{s(p)}{s(x)}$, the logarithm of the factor by which $s(x)$ grows, and these terms sum to at most $\log_2 n$. Case (A) contributes at most $2\log_{3/2}n+2\log_2n$.

\emph{Case (B).} If the skip of $p$ is heavy, the iteration contributes at most $1-2<0$. Otherwise $s_{i+1}>2s_i$ for the pair $(x_i,x_{i+1})=(p,g)$. The pairs are disjoint and the sizes increase along the path, so this happens at most $\log_2 n$ times, and $\Phi$ does not increase.

\emph{Case (C).} If the rotation of $p$ is heavy, the iteration contributes at most $1-2<0$. Otherwise $s_{i+1}>\frac32 s_i$, which happens at most $\log_{3/2}n$ times, and the increases $\log_2\frac{s_{i+1}}{s_i}$ over disjoint pairs sum to at most $\log_2 n$. Case (C) contributes at most $\log_{3/2}n+2\log_2n$.

Summing gives $W+2\Delta\Phi\le3\log_{3/2}n+5\log_2n$, and $3/\log_2(3/2)+5<10.13$.
\end{proof}

\begin{corollary}\label{cor:total}
Any sequence of $m$ operations performs at most $20.3\,m\log_2n+2\log_2n$ loop iterations. Hence $\code{AddEdge}$ and $\code{Query}$ run in $O(\log n)$ amortized time.
\end{corollary}
\begin{proof}
Each operation performs two accesses. Sum \cref{thm:main} over all $2m$ accesses, and use $\Phi_{\mathrm{initial}}=\log_2n$ and $\Phi_{\mathrm{final}}\ge0$.
\end{proof}

\begin{remark}\label{rem:necessary}
Both repairs of $p$ are needed. Deleting the lines of cases (B) and (C) from \cref{fig:code} leaves the naive loop, in which $x$ only rotates or skips; this is the promotion of the AM-tree without its size-based rebalancing. In our experiments, two inputs show that each repair is needed on its own. On the star built by $\code{AddEdge}(v,0,v)$ for $v=1,\dots,n-1$, accessing $1,2,\dots,n-1$ in order costs about two iterations per access with the loop of \cref{fig:code}, but about $n/2$ without case (C). On the path built by $\code{AddEdge}(v,v-1,n-v)$ for $v=1,\dots,n-1$, whose costs increase towards vertex~$0$, the queries $\code{Query}(i,i+1)$ for $i=0,1,\dots$ cost about four iterations each with the loop of \cref{fig:code}, but about $n/2$ without case (B). Without case (B), the line of case (C) must also test $\cost[p]\le\cost[g]$, so that $p$ rotates only at an up edge.
\end{remark}

\section{The structure as an implicit link-cut tree}\label{sec:lct}

This section interprets the structure; the analysis of \cref{sec:analysis} does not depend on it. Here, $G$ is connected and all inserted costs are distinct. Then $T$ contains no placeholder edges, and its costs are distinct as well: for every $t$, the number of edges of cost $t$ in a tree is determined by its bottleneck distances, so by Invariant~(I) it is the same for $T$ as for the MSF. The \emph{Kruskal reconstruction tree} $\KRT$ of a spanning tree, also known as its single-linkage dendrogram or the Cartesian tree of its edges~\cite{DemaineLW14}, has the vertices as leaves and one internal node per edge, created in increasing order of cost: the node of an edge has the two components joined by the edge as children, and its \emph{level} is the edge cost. By Invariant~(I), the reconstruction tree of $T$ equals that of the MSF of $G$. Since rotations and skips preserve all bottleneck distances, accesses do not change $\KRT$.

\begin{definition}
A non-root vertex $v$ \emph{owns} the internal node of $\KRT$ at level $\cost[v]$. For an internal node $Y$, the \emph{representative} $\mathrm{rep}(Y)$ is the unique leaf of $Y$ that owns no node inside $Y$. The \emph{preferred child} of $Y$ is the child that contains $\mathrm{rep}(Y)$.
\end{definition}

\begin{figure}[t]
\centering
\begin{tikzpicture}[v/.style={circle,draw,minimum size=6mm,inner sep=0pt}, n/.style={rectangle,draw,rounded corners=2pt,minimum size=5.5mm,inner sep=1pt,font=\small}, lab/.style={font=\scriptsize}, pref/.style={line width=2pt,blue!55!black}, >=stealth]
\begin{scope}
\node[v] (t4) at (0,4) {$4$}; \node[v] (t3) at (0,3) {$3$}; \node[v] (t2) at (0,2) {$2$}; \node[v] (t1) at (0,1) {$1$}; \node[v] (t0) at (0,0) {$0$};
\draw (t4)--node[lab,right]{$4$}(t3); \draw (t3)--node[lab,right]{$2$}(t2); \draw (t2)--node[lab,right]{$3$}(t1); \draw (t1)--node[lab,right]{$1$}(t0);
\node at (0,-0.8) {tree $T$};
\end{scope}
\begin{scope}[xshift=3.2cm]
\node[n] (d) at (2.6,4) {$4$}; \node[n] (c) at (1.5,3) {$3$}; \node[n] (a) at (0.6,2) {$1$}; \node[n] (b) at (2.4,2) {$2$};
\node[v] (l0) at (0,0.8) {$0$}; \node[v] (l1) at (1.1,0.8) {$1$}; \node[v] (l2) at (1.9,0.8) {$2$}; \node[v] (l3) at (2.9,0.8) {$3$}; \node[v] (l4) at (3.8,3) {$4$};
\draw (d)--(c); \draw[pref] (d)--(l4); \draw (c)--(a); \draw[pref] (c)--(b); \draw (a)--(l0); \draw[pref] (a)--(l1); \draw (b)--(l2); \draw[pref] (b)--(l3);
\node[lab,right] at (d.east) {owner $3$}; \node[lab,left] at (c.west) {owner $1$}; \node[lab,left] at (a.west) {owner $0$}; \node[lab,right] at (b.east) {owner $2$};
\node at (1.9,-0.1) {reconstruction tree $\KRT$};
\end{scope}
\end{tikzpicture}
\caption{A tree $T$ with $\link$ and $\cost$ (left) and its Kruskal reconstruction tree with internal nodes labeled by level (right). Thick edges lead to preferred children. The path ending at leaf~$1$ consists of the node of level~$1$; its parent node, of level~$3$, prefers the subtree $\{2,3\}$ with representative~$3$. The pointer $\link[1]=2$ is stale, and correspondingly the edge of vertex~$1$ is a down edge ($3>2$).}
\label{fig:krt}
\end{figure}
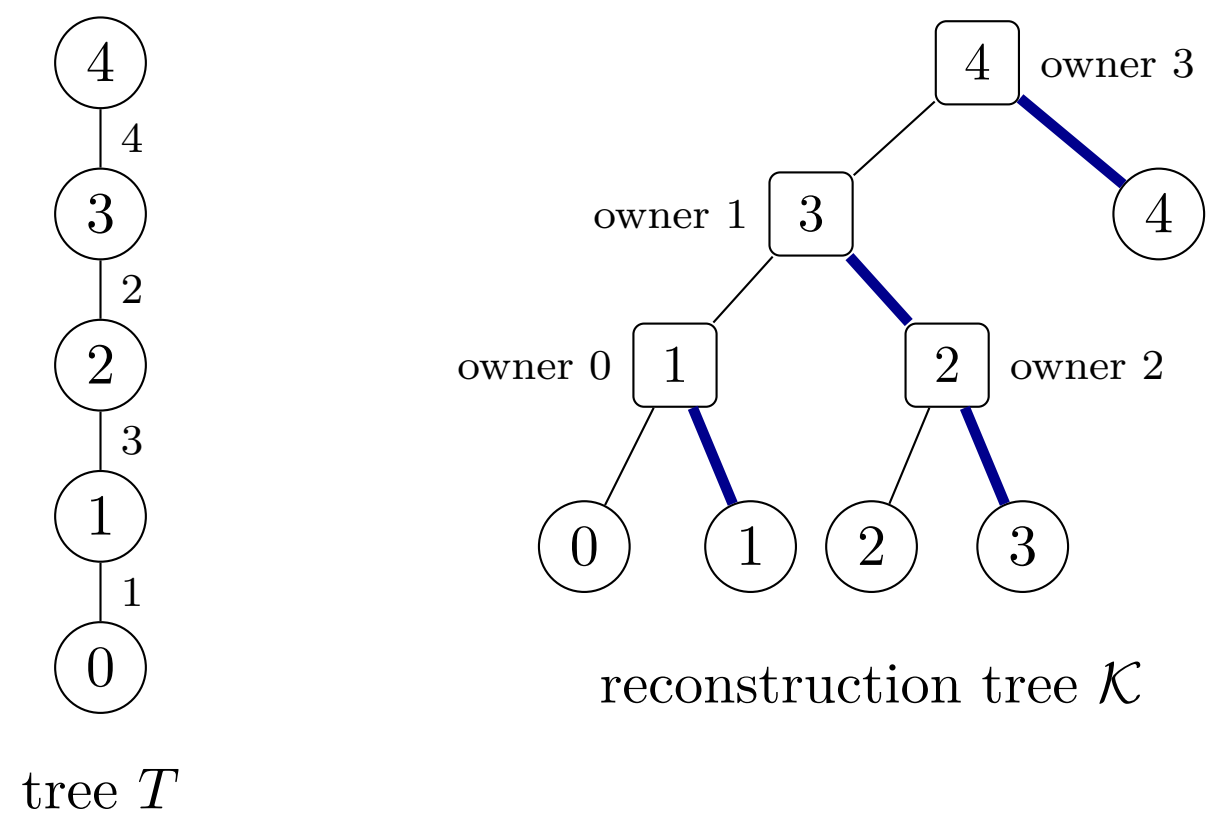

The representative is well defined. The node created by the edge of $v$ contains $v$, so each of the $k-1$ internal nodes of a node with $k$ leaves is owned by a distinct leaf of that node, and exactly one leaf owns none of them. As in a link-cut tree, preferred children partition $\KRT$ into paths that end at leaves. For a non-root vertex $v$, let $A_v$ be the node it owns. We call $\link[v]$ \emph{stale} if it is not $\mathrm{rep}(A_v)$. \Cref{fig:krt} shows an example.

\begin{lemma}\label{lem:stale}
Let $v$ be a non-root vertex and $p=\link[v]$.
\begin{enumerate}[label=(\roman*),leftmargin=2em]
\item The preferred path ending at $v$ consists of the nodes below $A_v$ that contain $v$, and $p$ lies in the preferred child of $A_v$. Thus $\link[v]$ is a path-parent pointer: it names a leaf of the preferred subtree of the parent node $A_v$, though not necessarily its representative.
\item $\link[v]$ is stale if and only if the edge of $v$ is a down edge.
\end{enumerate}
\end{lemma}
\begin{proof}
(i) A node $Y$ containing $v$ has $\mathrm{rep}(Y)=v$ exactly when $v$ owns no node inside $Y$, that is, when $Y$ lies strictly below $A_v$. Let $C$ be the child of $A_v$ that contains $v$. Its internal nodes are owned by its leaves other than $v$, one each, so every leaf of $C$ owns a node inside $A_v$, and $\mathrm{rep}(A_v)$ lies in the other child. That child also contains $p$, because $A_v$ is the node of the edge $\{v,p\}$ and joins the components of $v$ and $p$.

(ii) If $\cost[p]<\cost[v]$, then $A_p$ contains $p$ and lies below $A_v$, so $p$ owns a node inside $A_v$ and is not its representative. Otherwise $p$ is the root or $\cost[p]>\cost[v]$, so $p$ owns no node inside $A_v$, and $p=\mathrm{rep}(A_v)$ by uniqueness.
\end{proof}

\begin{theorem}\label{thm:lct}
Let $G$ be connected with distinct edge costs.
\begin{enumerate}[label=(\roman*),leftmargin=2em]
\item A skip changes no preferred child.
\item A rotation of $x$ changes exactly one preferred child: that of the node owned by $x$ before the rotation, which turns towards $x$.
\item After $\code{access}(z)$, every node on the path from $z$ to the root of $\KRT$ prefers the child containing $z$. The number of case-(A) rotations equals the number of nodes on this path that did not prefer $z$'s side before the access, that is, the number of preferred-child changes in a link-cut tree access to $z$ on $\KRT$. Case-(C) rotations change preferred children only off this path.
\end{enumerate}
\end{theorem}
\begin{proof}
Ownership, and hence every representative and preferred child, is determined by the costs alone.

(i) A skip changes no cost.

(ii) Let $A$ be the node of the edge $\{x,p\}$, at level $c_x$. If $p$ is not the root, let $A'\supset A$ be the node of the edge $\{p,g\}$, at level $c_p>c_x$. Before the rotation, $x$ owns $A$ and $p$ owns $A'$, or nothing if $p$ is the root. After the rotation, the owners are exchanged, and no other ownership changes. Consider an internal node $Y$. If $Y$ does not contain $A$, then $Y$ contains neither $A$ nor $A'$, so its internal nodes and their owners are unchanged, and so is $\mathrm{rep}(Y)$. If $Y$ contains $A'$, the set of owners of its internal nodes is unchanged, and so is $\mathrm{rep}(Y)$. Otherwise $A\subseteq Y$ and $p$ owns nothing inside $Y$. Then $\mathrm{rep}(Y)=p$ before the rotation and $\mathrm{rep}(Y)=x$ after it. If $Y\neq A$, the leaves $x$ and $p$ lie in the same child of $Y$, so the preferred child of $Y$ is unchanged. If $Y=A$, they lie in different children, and the preferred child of $A$ turns towards $x$.

(iii) At the end of the access, $z$ is the root of $T$ and owns no node. Hence $z$ is the representative of every node that contains it. A case-(C) rotation turns the node owned by $p$, at level $\cost[p]<\cost[z]$. This node does not contain $z$, because every path from $z$ to $p$ uses the edge of $z$, whose cost exceeds the node's level. A case-(A) rotation turns the node owned by $z$, which contains $z$, towards $z$. So nodes on the path of $z$ change only by case-(A) rotations, and each changes at most once, towards $z$. Every node on the path that did not prefer $z$'s side therefore changes exactly once, and no other node on the path changes.
\end{proof}

\Cref{thm:lct} accounts for the efficiency of the structure. In a link-cut tree, every preferred-child change requires splaying in an auxiliary tree to find where the path-parent pointer attaches. Here, the change is a single rotation, because the pointer names a leaf rather than a position in an auxiliary tree. As a consequence, pointers become stale when the representative of a path changes, and by \cref{lem:stale} the algorithm sees a stale pointer as a down edge. A skip of $v$ replaces $\link[v]$ by another leaf of the same preferred subtree, since $\bd(v,g)=\cost[v]$. After a skip in case (D) the pointer is fresh, because the new parent has a larger cost; after a skip in case (B) or (C) it may still be stale. The skips thus shortcut stale pointers without changing $\KRT$ or its preferred children, in a way similar to path halving in union-find~\cite{TarjanvL84}. Insertions, unlike accesses, change $\KRT$, so the usual bound on preferred-child changes in link-cut trees does not transfer directly.

Informally, if all pointers were fresh, contracting each preferred path to its bottom leaf would turn $T$ into the virtual tree of a link-cut tree. Stale pointers attach a path to another leaf of the right subtree, so $T$ is a lazily maintained version of that tree, and $\Phi$ is an analogue of the rank potential of Sleator and Tarjan~\cite{SleatorT85}.

\section{The folklore structure}\label{sec:folklore}

\Cref{fig:folklore} shows the folklore \code{BottleneckTree}~\cite{BudauBicsi21}, adapted to \code{int} costs. It uses the same representation and moves but organizes an access top-down. Procedure $\code{compress}(x)$ first recurses to the root. On the way back, at each down edge, it rotates the parent of $x$ upwards over all ancestors whose cost does not exceed $\cost[x]$, and then lets $x$ skip once. Finally, $\code{reroot}(x)$ rotates $x$ to the root.

\begin{figure}[t]
\begin{lstlisting}
int AddEdge(int x, int y, int c) {
  compress(x); reroot(x); compress(y); reroot(y);
  assert(link[x] == y);
  if (cost[x] > c) swap(cost[x], c);
  return c;
}
int Query(int x, int y) {
  compress(x); reroot(x); compress(y); reroot(y);
  return cost[x];
}
void compress(int x) {
  if (link[x] == -1) return;
  compress(link[x]);
  reroot(link[x], cost[x]);
  if (cost[link[x]] <= cost[x]) link[x] = link[link[x]];
}
void reroot(int x, int c = INF + 1) {
  int p = link[x];
  while (p != -1 && cost[p] <= c) {
    swap(cost[p], cost[x]);
    int g = link[p]; link[p] = x; p = g;
  }
  link[x] = p;
}
\end{lstlisting}
\caption{The structure of~\cite{BudauBicsi21}, with the same fields and constructor as in \cref{fig:code}.}
\label{fig:folklore}
\end{figure}

We measure the work of the folklore \code{BottleneckTree} as the number of calls of \code{compress} plus the number of iterations of the loop in \code{reroot}; each takes constant time.

\begin{theorem}\label{thm:folklore}
Let $L\ge4$ be even and $n>L^2/4+L/2+2$. Insert the star $\code{AddEdge}(v,0,v)$ for $v=1,\dots,n-1$. Then perform queries on the following sequence of vertices, taken in consecutive pairs. First, for $r=L-1,L-2,\dots,0$, the vertices congruent to $r$ modulo $L$ in decreasing order. Then $1,2,\dots,n-1$. On these $n-1$ queries, the folklore \code{BottleneckTree} performs at least
\[
\Bigl(\tfrac52L-1\Bigr)\Bigl(n-\tfrac{L^2}4-\tfrac L2-2\Bigr)
\]
work. In particular, for $L=2\lfloor\sqrt{n/3}\rfloor$ its amortized work per query is at least $1.9\sqrt n$, and per operation, including the $n-1$ insertions, at least $0.95\sqrt n$, for all sufficiently large $n$.
\end{theorem}

The choice $L\approx2\sqrt{n/3}\approx1.15\sqrt n$ maximizes the bound; our experiments use $L\approx1.1\sqrt n$. We prove \cref{thm:folklore} in \cref{app:folklore}, and sketch the mechanism here. After the first phase, every vertex $v\ge L$ points to $v-L$. In the second phase, the structure passes through states that repeat up to a shift by one position. Just before vertex $x$ is accessed, the vertices below $x$ form groups of $L-2,L-4,\dots,2$ consecutive vertices with stale pointers, separated by pairs of vertices with up edges, and the vertices above $x$ still point $L$ positions down. The root path of $x$ enters each group at its lowest vertex, so it has length $3L/2$, and the access performs a rollback of two rotations at each of the $L/2-1$ groups. Each group then loses its lowest vertex to the pair below it and gains a vertex from the pair above it, which restores the pattern one position higher.

Informally, in terms of $\Phi$, each rollback rotates a vertex with a small subtree over ancestors with much larger subtrees, at constant cost. This increases $\Phi$ by up to the logarithm of the ratio of their subtree sizes, and the input above causes $L/2-1$ such rollbacks per access. The rotations of the folklore \code{BottleneckTree} also involve a vertex and its parent, so this is not where the two structures differ. The difference is which parent. In the structure of \cref{fig:code}, the vertex rotated over is an original ancestor with its original subtree size (\cref{lem:untouched}), so the increases of $\Phi$ telescope along the root path. In the folklore \code{BottleneckTree}, $\code{compress}(\link[x])$ has already restructured the ancestors of $x$ when the rollback at $x$ begins, so this telescoping argument does not apply.

\section{Experiments}\label{sec:experiments}

\paragraph{Implementations.} We compare seven implementations:
\begin{itemize}[leftmargin=1.5em]
\item the structure of \cref{fig:code};
\item the folklore \code{BottleneckTree} (\cref{fig:folklore});
\item a textbook link-cut tree~\cite{SleatorT85} with edges represented as nodes, splay-based access, and evert by reversal bits;
\item the public AM-tree implementation~\cite{DingGS25,AMtreeCode} in its four variants: \emph{lazy} and \emph{strict}, each with \emph{perch} and with \emph{stitch} linking.
\end{itemize}
The AM-tree code uses $-1$ as the stored weight of a root, and its own driver inserts negative weights; we therefore map costs in $[0,10^9)$ order-preservingly to negative integers.

\paragraph{Workloads.} All workloads are on $n=2^{20}$ vertices. The first three consist of $4n$ insertions with costs drawn uniformly from $[0,10^9)$; the vertex labels are randomly permuted and the edges are inserted in random order.
\begin{itemize}[leftmargin=1.5em]
\item \emph{Random}: the endpoints are drawn uniformly from $V$. The final MSF has diameter about $870$.
\item \emph{Grid}: the vertices are placed on a $\sqrt n\times\sqrt n$ grid; for each vertex $v$, $4$ edges join a uniform vertex within distance $2$ of $v$ to one within distance $5$ of $v$ (in each coordinate, clamped to the grid). Diameter about $2\,500$.
\item \emph{Band}: for $i=0,\dots,n-1$, $4$ edges join a uniform $x\in[i-10,i]$ to a uniform $y\in[i,i+10]$ (clamped to $V$). Diameter about $166\,000$.
\end{itemize}
In \emph{Grid-local} and \emph{Band-local}, the same edges are inserted in their natural order, row by row and by increasing $i$, so that consecutive insertions are close in the graph; the labels remain permuted, so they are not close in memory.
\begin{itemize}[leftmargin=1.5em]
\item \emph{Lag}: the input of \cref{thm:folklore} with $L=2\lfloor0.55\sqrt n\rfloor$: a star of $n-1$ insertions followed by $n-1$ queries, adversarial for the folklore \code{BottleneckTree}.
\end{itemize}
For the folklore \code{BottleneckTree} we also measure the query phase of Lag for $n=2^{12},2^{14},\dots,2^{20}$.

\paragraph{Correctness.} All implementations returned identical answers on all workloads, including 200\,000 random queries after each run, which also check the insertion-only workloads. Against a brute-force minimax computation, the structure of \cref{fig:code} returned the correct value on 517\,000 random operations with frequent equal costs. For $n\le5$ we enumerated every reachable state with every possible \code{AddEdge}, 9.1 million operations including ties; every answer was correct. Over 2.2 million accesses on random and adversarial inputs, every inequality used in the proof of \cref{thm:main} held, and the largest value of $(W+2\Delta\Phi)/\log_2n$ was $4.1$, compared with the bound $10.13$ of \cref{thm:main}. The statements of \cref{thm:lct} were checked on 297\,000 accesses.

\paragraph{Setup.} All code was compiled with \code{g++} 13.3 using \code{-O3 -march=native} on Ubuntu 24.04 and run single-threaded, pinned to one core, on two Amazon EC2 instances: \code{c7a.2xlarge} (AMD EPYC 9R14) and \code{c7i.12xlarge} (Intel Xeon Platinum 8488C). All structures use 32-bit costs. Each implementation is compiled into a separate program, because we observed that other code in the same program can change how a structure's main loop is compiled, by up to 50\% in running time. Each entry of \cref{tab:time} is the median of five runs in a fresh process. To exclude drift, we also ran our structure and the lazy AM-tree variants in 11 interleaved rounds; the median per-round ratios agreed with \cref{tab:time} within 7\% on the insertion workloads and within 20\% on Lag, whose runs take 25\,ms.
\begin{table}[t]
\centering\footnotesize\setlength{\tabcolsep}{2.4pt}
\begin{tabular}{@{}l rr rrrr rrrr rr@{}}
\toprule
& \multicolumn{2}{c}{\multirow{2}{*}{Random}} & \multicolumn{4}{c}{Grid} & \multicolumn{4}{c}{Band} & \multicolumn{2}{c}{\multirow{2}{*}{Lag}} \\
\cmidrule(lr){4-7}\cmidrule(lr){8-11}
& \multicolumn{2}{c}{} & \multicolumn{2}{c}{random} & \multicolumn{2}{c}{local} & \multicolumn{2}{c}{random} & \multicolumn{2}{c}{local} & \multicolumn{2}{c}{} \\
\cmidrule(lr){2-3}\cmidrule(lr){4-5}\cmidrule(lr){6-7}\cmidrule(lr){8-9}\cmidrule(lr){10-11}\cmidrule(l){12-13}
& AMD & Intel & AMD & Intel & AMD & Intel & AMD & Intel & AMD & Intel & AMD & Intel \\
\midrule
ours & 1065 & 1695 & 1245 & 1934 & 367 & 477 & 1198 & 1935 & 198 & 271 & 24.9 & 27.3 \\
folklore & 2268 & 3155 & 2214 & 3254 & 618 & 713 & 1846 & 2593 & 299 & 357 & 10642 & 12975 \\
LCT & 12748 & 13790 & 11172 & 12736 & 3579 & 3423 & 10417 & 12097 & 2136 & 1996 & 398 & 319 \\
\addlinespace
AM-LP & 776 & 1350 & 1012 & 1759 & 385 & 493 & 1329 & 2322 & 245 & 302 & 25.1 & 25.0 \\
AM-LS & 624 & 1053 & 655 & 1012 & 283 & 369 & 641 & 975 & 318 & 442 & 25.5 & 25.6 \\
AM-SP & 6856 & 7975 & 7872 & 9258 & 2182 & 2292 & 10129 & 12263 & 2012 & 2296 & 29.4 & 31.0 \\
AM-SS & 2446 & 2914 & 2500 & 3077 & 755 & 898 & 2698 & 3227 & 812 & 947 & 31.3 & 31.4 \\
\bottomrule
\end{tabular}
\caption{Running time in milliseconds on the AMD and Intel machines (median of five runs; each entry measured in a fresh process). Random has its edges inserted in random order, Grid and Band in random and in local order. Lag is the adversarial input of \cref{thm:folklore}. The row \emph{folklore} is the folklore \code{BottleneckTree}, LCT the link-cut tree; AM-LP and AM-LS are the lazy AM-tree with perch and with stitch linking, AM-SP and AM-SS the strict one. All implementations returned identical answers.}
\label{tab:time}
\end{table}

\paragraph{Results.} On the insertion workloads, the folklore \code{BottleneckTree} is 1.3--2.1 times slower than our structure, and the link-cut tree is 6.3--12 times slower (\cref{tab:time}). The strict AM-tree variants, which also bound the worst-case query time, are slower than ours throughout. Against the lazy variants, the outcome depends on the order of insertions. In random order, stitch linking is 1.6--2.0 times faster than ours on all three random-order workloads, and perch linking is 1.3--1.4 times faster on Random and 1.1--1.2 times faster on Grid, but 1.1--1.2 times slower on Band. In local order, perch linking is 3--5\% slower than ours on Grid-local, and stitch linking 1.3 times faster; on Band-local, our structure is the fastest of all, and the lazy variants are 1.1--1.2 times slower with perch and 1.6 times slower with stitch linking. On Lag, the folklore \code{BottleneckTree} is 430--480 times slower than our structure and the link-cut tree 12--16 times slower, while the AM-tree variants take 0.92--1.3 times our time. From $n=2^{12}$ to $2^{20}$, the folklore time per query on Lag grows by a factor of about 22, somewhat more than $\sqrt n$ (a factor of 16), while ours changes by at most 18\%.

\begin{table}[!ht]
\centering\small
\begin{tabular}{@{}l r rr rr@{}}
\toprule
& \multirow{2}{*}{Random} & \multicolumn{2}{c}{Grid} & \multicolumn{2}{c}{Band} \\
\cmidrule(lr){3-4}\cmidrule(l){5-6}
& & random & local & random & local \\
\midrule
ours & 22.2 & 25.5 & 10.9 & 24.6 & 6.6 \\
AM-LP & 9.7 & 13.0 & 6.6 & 20.2 & 5.1 \\
AM-LS & 6.8 & 8.3 & 4.0 & 11.0 & 10.3 \\
\bottomrule
\end{tabular}
\caption{Average depth walked per insertion: the sum of the depths of the two endpoints just before they are accessed, which is the number of vertices on the root paths that the insertion follows. Measured on instrumented copies of the code, on the inputs of \cref{tab:time}.}
\label{tab:ops}
\end{table}

\Cref{tab:ops} explains these results. The running time follows the number of vertices on the root paths that an insertion visits, since each is a dependent memory access: the next vertex is known only after the current one has been read. Our structure reads two arrays per vertex and the AM-tree three, so each vertex costs us less. In random order, our paths are 1.2--3.3 times longer than those of the lazy AM-tree, whose size rule keeps the tree shallow. In local order, the endpoints of a new edge are close to vertices that recent accesses brought near the root, and our paths become 2.3--3.7 times shorter (10.9 and 6.6 vertices). Those of perch linking shrink about as much, and those of stitch linking do not shrink on Band.

The trade-off resembles the one between splay trees~\cite{SleatorT85} and weight-balanced search trees~\cite{NievergeltR73}. Like a splay tree, our structure stores no balance information and restructures along every access path. The anti-monopoly rule of the AM-tree is a weight-balance condition with parameter $2/3$, which the lazy variant restores only where it is violated. In experimental comparisons of search trees, splay trees are likewise slower than balanced trees on uniformly random accesses and faster on accesses with locality~\cite{Pfaff04}, such as sequential access, on which they take linear total time~\cite{Tarjan85}.

\section{Related work}\label{sec:related}

Link-cut trees~\cite{SleatorT83,SleatorT85} support incremental MST in $O(\log n)$ amortized time; top trees~\cite{AlstrupHLT05,TarjanW05} and RC-trees~\cite{AcarBV05} give alternatives, and Tarjan and Werneck~\cite{TarjanW09} compare these structures experimentally, including on online MSF. The fully dynamic problem, with deletions, requires substantially more machinery~\cite{Frederickson85,HolmLT01}, and offline variants have simpler solutions~\cite{Eppstein94}. Batch-incremental and sliding-window variants were studied by Anderson, Blelloch and Tangwongsan~\cite{AndersonBT20}.

The D-tree~\cite{ChenHLB22} maintains rooted spanning trees for fully dynamic connectivity queries. Closer to our setting, the OEC-forest~\cite{SongWXQZL24} maintains a rooted tree with the path maxima of the MSF and uses the transformation behind \cref{lem:moves} for temporal connectivity. The AM-tree~\cite{DingGS25} added the anti-monopoly rule and obtained $O(\log n)$ amortized bounds, with worst-case query bounds in its strict variant. As discussed in the introduction, our structure differs from it mainly in the order of operations within an access, and it stores no sizes. The AM-tree offers things ours does not: its queries are read-only once sizes are calibrated, its strict variant bounds the worst-case query time, and it has a persistent version for historical queries on temporal graphs. Because every access of our structure restructures the tree, persistence would be costly for it. In the Kruskal reconstruction tree, also known as the Cartesian tree of the MSF edges~\cite{DemaineLW14}, an insertion merges two root paths, as in mergeable trees~\cite{GeorgiadisKSTW11}. The advancing of stale pointers resembles path halving in union-find~\cite{TarjanvL84} and union-find with deletions~\cite{KaplanST02}.

The representation also appears in competitive programming. A January 2020 solution by Basu~\cite{Basu20} to an incremental MST problem~\cite{Rutier} merges the root paths of the two endpoints of a new edge, guided by subtree sizes, in much the same way as the stitch linking of the AM-tree, but without rebalancing and without a bound. The folklore \code{BottleneckTree} grew out of it, replacing the sizes by rotations. A structure for the same problem with a claimed $O(\log n)$ bound was described in a Codeforces blog post~\cite{NotAmir24}. It does not come with a proof.

\section{Conclusion and open problems}\label{sec:conclusion}

The structure of \cref{fig:code} solves incremental MST and bottleneck queries in $O(\log n)$ amortized time with two arrays and one loop. In practice it is 1.3--2.1 times faster than the folklore \code{BottleneckTree} on insertions and 6.3--16 times faster than an explicit link-cut tree; the lazy AM-tree variants are faster when edges arrive in random order, and ours can be faster when consecutive insertions are close in the graph (\cref{sec:experiments}). Its analysis follows from one observation: during an access, the ancestors of the current vertex are untouched, so every rotation is over an original ancestor with its original subtree size, and the potential gains telescope.

\paragraph{Optimality.} The bound is optimal for structures that only compare costs. Insert $n$ keys as the edges of a path, then insert $n$ edges of cost $-\infty$ (any cost below all keys), each between the two current endpoints of the path. Each insertion closes a cycle through all remaining keys and evicts the largest, whose endpoints become the new endpoints. The evicted costs come out in decreasing order, so $2n$ operations sort $n$ keys, and some operation takes $\Omega(\log n)$ amortized time. In the cell-probe model, the $\Omega(\log n)$ bound of P\v{a}tra\c{s}cu and Demaine~\cite{PatrascuD06} for dynamic connectivity gives a related bound for dynamic MSF with deletions.

\paragraph{Open problems.}
\begin{itemize}[leftmargin=1.5em]
\item Our bounds are amortized. Is there a comparably simple structure with $O(\log n)$ worst-case queries, as in the strict AM-tree?
\item With edges in random order, the root paths followed by our structure are up to 3.3 times longer than those of the lazy AM-tree (\cref{tab:ops}). Is there an equally simple structure without sizes that keeps paths as short on such inputs?
\item By \cref{thm:lct}, the case-(A) rotations of an access are the preferred-child changes on the Kruskal reconstruction tree, the quantity behind Wilber's first lower bound~\cite{Wilber89} and Tango trees~\cite{DemaineHIP07}. Does the structure satisfy analogues of the splay-tree theorems, such as sequential access or working set, and is there a meaningful notion of dynamic optimality for path-maxima trees?
\end{itemize}

\section*{Acknowledgments}
We thank Rajarshi Basu, whose January 2020 solution to the problem \emph{Transport Rutier}~\cite{Rutier} on the infoarena judge contained the first version of this structure; the question of its complexity started this line of work. We used an AI assistant (Claude) for experiments, verification scripts and editing; all results were checked by the authors.

\appendix
\section{Proof of \texorpdfstring{\cref{thm:folklore}}{the lower bound}}\label{app:folklore}

\begin{figure}[t]
\centering
\begin{tikzpicture}[b/.style={rectangle,draw,fill=black!18,minimum size=4.6mm,inner sep=0pt,font=\scriptsize}, o/.style={circle,draw,minimum size=4.6mm,inner sep=0pt,font=\scriptsize}, pt/.style={->,>=stealth,black!55}, hl/.style={->,>=stealth,very thick,blue!60!black}]
\node[b] (na_30) at (0.00,2.2) {30};
\node[b] (na_31) at (0.72,2.2) {31};
\node[b] (na_32) at (1.44,2.2) {32};
\node[b] (na_33) at (2.16,2.2) {33};
\node[o] (na_34) at (2.88,2.2) {34};
\node[o] (na_35) at (3.60,2.2) {35};
\node[b] (na_36) at (4.32,2.2) {36};
\node[b] (na_37) at (5.04,2.2) {37};
\node[o] (na_38) at (5.76,2.2) {38};
\node[o] (na_39) at (6.48,2.2) {39};
\node[o] (na_40) at (7.20,2.2) {40};
\node[o] (na_41) at (7.92,2.2) {41};
\node[b] (na_42) at (8.64,2.2) {42};
\node[b] (na_43) at (9.36,2.2) {43};
\node[o] (na_44) at (10.08,2.2) {44};
\node[o] (na_45) at (10.80,2.2) {45};
\node[o] (na_46) at (11.52,2.2) {46};
\node[o] (na_47) at (12.24,2.2) {47};
\draw[hl] (na_34) to[bend right=50] (na_32);
\draw[pt] (na_35) to[bend right=50] (na_33);
\draw[hl] (na_38) to[bend right=50] (na_34);
\draw[pt] (na_39) to[bend right=50] (na_35);
\draw[pt] (na_40) to[bend right=50] (na_36);
\draw[pt] (na_41) to[bend right=50] (na_37);
\draw[hl] (na_44) to[bend right=50] (na_38);
\draw[pt] (na_45) to[bend right=50] (na_39);
\draw[pt] (na_46) to[bend right=50] (na_40);
\draw[pt] (na_47) to[bend right=50] (na_41);
\node[b] (nb_30) at (0.00,0) {30};
\node[b] (nb_31) at (0.72,0) {31};
\node[b] (nb_32) at (1.44,0) {32};
\node[b] (nb_33) at (2.16,0) {33};
\node[b] (nb_34) at (2.88,0) {34};
\node[o] (nb_35) at (3.60,0) {35};
\node[o] (nb_36) at (4.32,0) {36};
\node[b] (nb_37) at (5.04,0) {37};
\node[b] (nb_38) at (5.76,0) {38};
\node[o] (nb_39) at (6.48,0) {39};
\node[o] (nb_40) at (7.20,0) {40};
\node[o] (nb_41) at (7.92,0) {41};
\node[o] (nb_42) at (8.64,0) {42};
\node[b] (nb_43) at (9.36,0) {43};
\node[b] (nb_44) at (10.08,0) {44};
\node[o] (nb_45) at (10.80,0) {45};
\node[o] (nb_46) at (11.52,0) {46};
\node[o] (nb_47) at (12.24,0) {47};
\draw[pt] (nb_35) to[bend right=50] (nb_33);
\draw[pt] (nb_36) to[bend right=50] (nb_34);
\draw[pt] (nb_39) to[bend right=50] (nb_35);
\draw[pt] (nb_40) to[bend right=50] (nb_36);
\draw[pt] (nb_41) to[bend right=50] (nb_37);
\draw[pt] (nb_42) to[bend right=50] (nb_38);
\draw[pt] (nb_45) to[bend right=50] (nb_39);
\draw[pt] (nb_46) to[bend right=50] (nb_40);
\draw[pt] (nb_47) to[bend right=50] (nb_41);
\node[anchor=east,font=\small] at (-0.5,2.2) {$P(44)$};
\node[anchor=east,font=\small] at (-0.5,0) {$P(45)$};
\end{tikzpicture}
\caption{The steady pattern for $L=6$ ($K=2$), drawn on the vertices $30,\dots,47$; it is exactly the state of the folklore \code{BottleneckTree} before the accesses of $44$ and $45$. Squares are the elements of $B$, which form a chain of up edges; a circle $v$ has cost $v$ and an arrow to its pointer $p_v$. In $P(44)$, the groups $H_1=\{38,\dots,41\}$ and $H_2=\{34,35\}$ point $4$ and $2$ positions down, the pairs $\{42,43\}$, $\{36,37\}$ and everything up to $33$ are in $B$, and the vertices from $44$ on point $6$ down. The access of $44$ follows the thick arrows $44\to38\to34\to32$ and rolls $34$ and $38$ back into $B$; the result is the same pattern one position higher.}
\label{fig:pattern}
\end{figure}
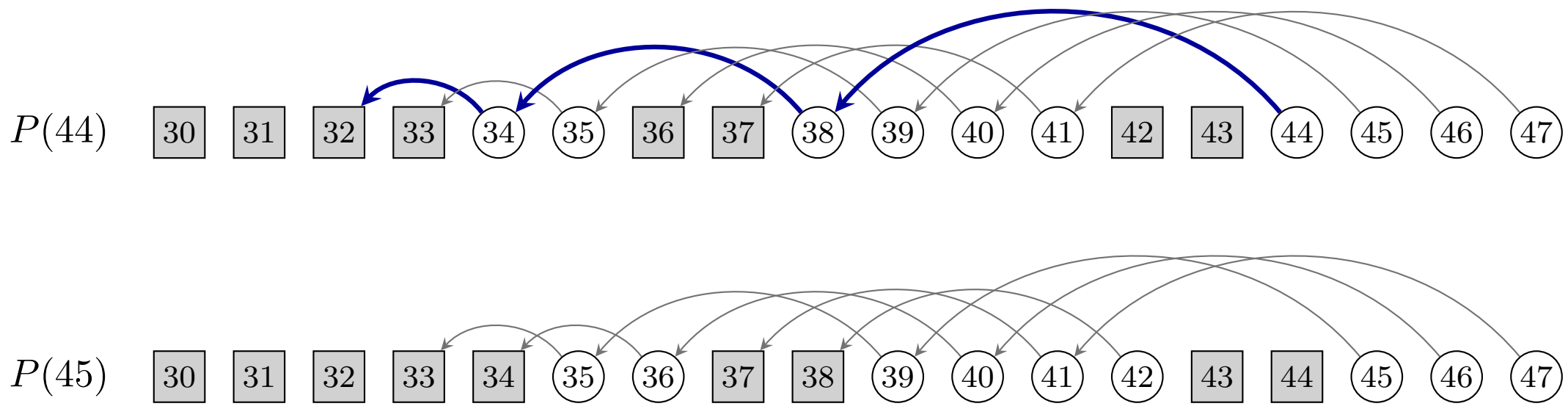

The proof follows the state of the folklore \code{BottleneckTree} exactly, access by access; a query performs the access of its first vertex and then the access of its second, so it suffices to follow accesses. It has four steps.
\begin{enumerate}[label=(\arabic*),leftmargin=2em]
\item The state before every access has a simple shape, which we call \emph{chain form}: a sorted chain $B$ of vertices ending at the root, and every other vertex pointing to a smaller vertex.
\item One access transforms a chain form by an explicit rule (\cref{lem:chainaccess}).
\item The first phase of queries produces a specific chain form (\cref{lem:firstphase}).
\item After a short warm-up of the second phase, the state before the access of $x$ is a fixed pattern $P(x)$, cut off at $0$, and the access turns it into the same pattern shifted by one position, $P(x+1)$ (\cref{lem:pattern}). For large enough $x$, each such access costs exactly $\frac52L-1$. \Cref{fig:pattern} shows the pattern for $L=6$.
\end{enumerate}
Throughout, $\mathrm{pred}_B(v)$ and $\mathrm{succ}_B(v)$ denote the largest element of a set $B$ below $v$ and the smallest element above $v$.

\paragraph{Chain form.} A state is in chain form if there is a set $B\subseteq V$ with $0\in B$, and every vertex $v\notin B$ has a \emph{pointer} $p_v<v$, such that:
\begin{itemize}[leftmargin=1.5em]
\item the largest element of $B$ is the root, and every other $b\in B$ has $\link[b]=\cost[b]=\mathrm{succ}_B(b)$, so the elements of $B$ form a chain of up edges ending at the root;
\item every $v\notin B$ has $\link[v]=p_v$ and $\cost[v]=v$.
\end{itemize}
We write such a state as $(B,p)$. In \cref{fig:pattern}, the elements of $B$ are drawn as squares and the pointers as arrows.

\paragraph{One access.} In chain form, the root path of a vertex $z$ first follows pointers downwards until it reaches $B$, and then climbs the chain $B$. On the way back up from this descent, the folklore code performs a \emph{rollback} of a vertex $y$ whenever it can rotate $y$ over a run of chain elements; all of them except the lowest leave $B$ and point to $y$, and $y$ joins $B$. The following lemma states exactly which elements move.

\begin{lemma}[Access in chain form]\label{lem:chainaccess}
Let $(B,p)$ be in chain form and $z>\max B$. Let $z=i_0>i_1>\dots>i_t$ be the vertices outside $B$ reached from $z$ by following pointers, and let $r=p_{i_t}\in B$. Then $\code{compress}(z)$ followed by $\code{reroot}(z)$ transforms $(B,p)$ as follows. Process $j=t,t-1,\dots,0$ in this order, with $B$ the current set:
\begin{enumerate}[label=(\alph*),leftmargin=2em]
\item let $y=i_{j+1}$ (where $i_{t+1}=r$), the vertex that $i_j$ points to;
\item let $Q=\mathrm{pred}_B(i_j)$ if $j\ge1$, and $Q=\max B$ if $j=0$;
\item let $\ell=y$ if $y\in B$, and $\ell=\mathrm{pred}_B(y)$ otherwise;
\item if $\ell<Q$, every $b\in B$ with $\ell<b<Q$ leaves $B$ with $p_b=y$, and $y$ joins $B$;
\item in any case, set $p_{i_j}=Q$.
\end{enumerate}
Finally, $z$ joins $B$.
\end{lemma}
\begin{proof}
The root path of $z$ is $i_0,\dots,i_t$ followed by the elements of $B$ from $r$ upwards. Along that chain the costs increase, so for every chain vertex $c$, $\code{compress}(c)$ changes nothing: the loop of $\code{reroot}(\link[c],\cost[c])$ stops at once, and $c$ does not skip. The calls $\code{compress}(i_j)$ then return for $j=t,\dots,0$ in this order. We show by induction that each implements step $j$ of the rule and leaves $\link[i_j]=Q=\mathrm{pred}_B(i_j)$ and $\cost[i_j]=i_j$, with $i_0,\dots,i_{j-1}$ untouched.

Consider $\code{reroot}(y,i_j)$. By induction, $y$ is either $r\in B$, or it is outside $B$ with $\link[y]=\mathrm{pred}_B(y)$. In both cases the loop walks up the chain $B$, from $\mathrm{succ}_B(y)$ if $y\in B$ and from $\ell$ otherwise. A chain element $b$ has cost $\mathrm{succ}_B(b)$, which is at most $i_j$ exactly when $b<Q$; the root has cost $\infty^+$. Hence the loop rotates $y$ over the chain elements $b_1<\dots<b_s$ in $[\ell,Q)$, excluding $\ell$ itself if $y\in B$. Afterwards each $b_k$ is a child of $y$; $b_1$ has the former cost of $y$, $b_k$ the former cost of $b_{k-1}$, which is $b_k$, for $k\ge2$; and $y$ has link and cost $Q$. If $y\in B$, the former cost of $y$ is $\mathrm{succ}_B(y)=b_1$, so all $b_k$ leave $B$ with pointer $y$, and $y$ stays in $B$ with successor $Q$. If $y\notin B$ and $s\ge1$, then $b_1=\ell$ gets link and cost $y$ and stays in $B$ with successor $y$, the other $b_k$ leave $B$ with pointer $y$, and $y$ joins $B$ with successor $Q$. If $s=0$, nothing changes. This is the rule. In every case, $\link[y]=Q$ or $y=Q$ afterwards. If $y=Q$, then $\cost[y]=\mathrm{succ}_B(Q)>i_j$ and $i_j$ keeps $\link[i_j]=y=Q$. Otherwise $\cost[y]\le i_j$, and $i_j$ skips to $\link[y]=Q$. No element between $Q$ and $i_j$ was added, so $Q$ is still $\mathrm{pred}_B(i_j)$.

For $j=0$, the same argument with $Q=\max B$ applies, since every chain element below the root has cost at most $\max B<z$. Finally, $\code{reroot}(z)$ rotates $z$ over the root $\max B$, which gets link and cost $z$, so $z$ joins $B$ as the new root.
\end{proof}

\paragraph{The first phase.} The first phase accesses the residue classes modulo $L$ from the top down. Its effect is that every vertex $v\ge L$ ends up pointing exactly $L$ positions below itself.

\begin{lemma}[First phase]\label{lem:firstphase}
After the star and the first phase, the state is in chain form with $B=\{0\}$, $p_v=v-1$ for $1\le v<L$, and $p_v=v-L$ for $v\ge L$.
\end{lemma}
\begin{proof}
Each $\code{AddEdge}(v,0,v)$ rotates $v$ over the root $0$ and then $0$ back over $v$, and the swap gives $v$ the cost $v$. After the star, the state is in chain form with $B=\{0\}$ and $p_v=0$ for all $v$.

We claim that before the access of a vertex $y$ of class $r$, the state is in chain form with $B=\{0,c,w\}$, where $c=r+1$ is the last vertex accessed in class $r+1$ (absent if $r=L-1$) and $w=y+L$ is the vertex accessed before $y$ in class $r$ (absent if $y$ is the first). Vertices that have not been accessed have pointer $0$, and every accessed vertex other than $c$ and $w$ already has its final pointer.

Let $y$ be accessed. Its root path is $y,0,c,w$, omitting absent vertices, and $y\notin B$ has pointer $0$. If $c$ is absent or $c<y$, the call $\code{compress}(y)$ changes at most the pointer of $y$: if $c$ is present, $y$ skips from $0$ to $c$, since $\cost[0]=c\le y<\cost[c]$; otherwise $\cost[0]=w>y$ and nothing happens. Then $\code{reroot}(y)$ rotates $y$ over the remaining vertices of its root path. Afterwards $B=\{0,c,y\}$ (or $\{0,y\}$), and $w$ has cost $w$ and pointer $y=w-L$. If $y=r<c$, then $\code{compress}(y)$ changes nothing, since $\cost[0]=c>y$, and $\code{reroot}(y)$ rotates $y$ over $0$, $c$ and $w$. Afterwards $B=\{0,r\}$, $c$ has pointer $r=c-1$, and $w$ has pointer $w-L$. The vertex accessed next is the first of class $r-1$, so the claim holds with $c=r$. The last access, of $y=0\in B$, rotates $0$ over $c=1$ and $w=L$, which then point to $0=1-1=L-L$.
\end{proof}

\paragraph{The steady pattern.} Before the access of $x$ in the second phase, the vertices below $x$ form groups of $L-2,L-4,\dots,2$ consecutive vertices, whose pointers jump $L-2,L-4,\dots,2$ positions down; consecutive groups are separated by a pair of elements of $B$, the two vertices just below $x$ are in $B$, and below the smallest group every vertex is in $B$. Above $x$, every vertex still points $L$ positions down. Formally, let $K=L/2-1$. For an integer $x$, define $u_0=x-1$ and $u_{j+1}=u_j-L+2j$ for $0\le j<K$. The pattern $P(x)$ is the chain-form state on the integers in which
\[
B=\{v\le u_K\}\cup\bigcup_{0\le j<K}\{u_j-1,u_j\},
\]
the vertices of the group $H_j=[u_j+1,\,u_{j-1}-2]$ have pointer $p_v=v-(L-2j)$ for $1\le j\le K$, and every $v\ge x$ has pointer $p_v=v-L$. Each $H_j$ consists of $L-2j$ consecutive vertices. For example, for $L=6$ and $x=44$ we get $K=2$, $u_0=43$, $u_1=37$ and $u_2=33$, so $B$ contains everything up to $33$ and the pairs $\{36,37\}$ and $\{42,43\}$, and the groups are $H_1=\{38,\dots,41\}$ and $H_2=\{34,35\}$ (\cref{fig:pattern}). For $n$ vertices, let $\pi(B,p)$ be the chain form on $\{0,\dots,n-1\}$ with set $(B\cap[1,n))\cup\{0\}$ and pointers $\max(p_v,0)$; it cuts the infinite pattern off at $0$.

The next lemma says that one access shifts the pattern by one position. In the example, the access of $44$ follows the pointers $44\to38\to34\to32$. On the way back up, no element of $B$ moves at $34$; only the pointer of $34$ advances from $32$ to $33$. At $38$, the code rolls $34$ back over $33$ and $36$: $36$ leaves $B$ and points to $34$, which joins $B$. At $44$, it rolls $38$ back over $37$ and $42$: $42$ leaves $B$ and points to $38$, which joins $B$. Finally $44$ joins $B$. The result is $P(45)$.

\begin{lemma}[Shift by one]\label{lem:pattern}
For every $x$, applying the rule of \cref{lem:chainaccess} to $P(x)$ with $z=x$ gives $P(x+1)$, and the rule commutes with $\pi$.
\end{lemma}
\begin{proof}
Let $d_j=u_j+1$, the lowest vertex of $H_j$. The pointers lead from $x$ to $d_1,d_2,\dots,d_K$ and then to $d_K-2=u_K-1\in B$. At step $K$, $y=u_K-1$ and $Q=u_K$, so nothing is removed, and $d_K$ gets pointer $u_K$. At step $j<K$, $y=d_{j+1}$, $\ell=u_{j+1}$ and $Q=u_j$, so $u_j-1$ leaves $B$ with pointer $d_{j+1}$, and $d_{j+1}$ joins $B$. Finally $x$ joins $B$. The new values $u_j'=u_j+1$ satisfy the same recurrence from $u_0'=x$. The new pairs $\{u_j,d_j\}$ and $\{u_0,x\}$ are $\{u_j'-1,u_j'\}$. The new group $H_j'$ is $H_j$ without $d_j$ and with $u_{j-1}-1$ added, whose pointer $d_j$ is $L-2j$ below it. This is $P(x+1)$.

For the commutation, write $\bar v=\max(v,0)$. For $v\ge1$, we have $v\in\pi(B)$ if and only if $v\in B$, and $\mathrm{pred}_{\pi(B)}(v)=\overline{\mathrm{pred}_B(v)}$; also $\max\pi(B)=\overline{\max B}$. In $\pi(B,p)$, the pointers from $z$ follow $i_0,i_1,\dots$ while these are positive, and the first non-positive one is replaced by $0\in\pi(B)$. Consider step $j$ of the rule on $(B,p)$. Since $y<i_j$, we have $\ell\le Q$, and $\ell<Q$ holds if and only if $B$ has an element in $(y,i_j)$, in which case $y<Q$. If $i_j\le0$, then $y$, $Q$ and every vertex the step changes are non-positive, so $\pi$ does not see the step. If $i_j\ge1$, the same step on $\pi(B,p)$ has $y'=\bar y$, $Q'=\bar Q$ and, in both cases of (c), $\ell'=\bar\ell$. If $Q\le0$, the step on $(B,p)$ changes only non-positive vertices and sets $p_{i_j}=Q$, while the step on $\pi(B,p)$ removes nothing and sets $p_{i_j}=0=\bar Q$. If $Q\ge1$, then $\ell'<Q'$ if and only if $\ell<Q$; the removed set $\pi(B)\cap(\ell',Q)$ is the positive part of $B\cap(\ell,Q)$, and its new pointer $\bar y$ is the clamped pointer $y$; $\bar y$ belongs to $\pi(B)$ afterwards exactly when $y$ joins $B$ or $y\le0$; and $p_{i_j}=Q=Q'$. Finally, $z\ge1$ joins both sets. Hence applying the rule to $\pi(B,p)$ gives $\pi$ of the result on $(B,p)$. We also checked this mechanically (see below).
\end{proof}

\paragraph{Counting the work.} It remains to check that the second phase reaches the pattern, and to count the work of one access in the pattern: one call of $\code{compress}$ for each of the $3L/2$ vertices on the root path, two rotations for each of the $K$ rollbacks, and one final rotation in $\code{reroot}$.

\begin{proof}[Proof of \cref{thm:folklore}]
Before access $x$ of the second phase, the root is $x-1=\max B$, so \cref{lem:chainaccess} applies. From the state of \cref{lem:firstphase}, the accesses $x=1,\dots,L-1$ follow the pointer $x-1=\max B$ and only add $x$ to $B$; the access of $L$ follows $p_L=0$ and removes $1,\dots,L-2$, which get pointer $0$. The result is $B=\{0,L-1,L\}$, which is $\pi(P(L+1))$. By \cref{lem:pattern}, the state before access $x$ is $\pi(P(x))$ for all $x\ge L+1$.

From the recurrence, $u_K=x-L^2/4-L/2+1$. If $x\ge L^2/4+L/2+1$, then $u_K-1\ge1$, so $\pi$ does not affect the root path of $x$, which consists of $x$, $d_1,\dots,d_K$, the pair $\{u_K-1,u_K\}$, and the $K$ pairs above it: $3L/2$ vertices. By the proof of \cref{lem:pattern}, the access rotates over two vertices at each of the steps $j<K$, and once in $\code{reroot}(x)$. Its work is therefore $3L/2+2K+1=\frac52L-1$. The sequence has $2n-1$ vertices, so the queries cover the accesses up to $x=n-2$, and $n-L^2/4-L/2-2$ of them satisfy this condition.

For $L=2\lfloor\sqrt{n/3}\rfloor$ we have $L^2/4\le n/3$ and $L\ge2\sqrt{n/3}-2$, so the work per query is at least $(5\sqrt{n/3}-6)(\tfrac23-O(n^{-1/2}))=\frac{10}{3\sqrt3}\sqrt n-O(1)$, and $\frac{10}{3\sqrt3}>1.92$.
\end{proof}

We checked each lemma mechanically. \Cref{lem:chainaccess} was compared with the code on 300\,000 random chain-form states, and \cref{lem:firstphase} and \cref{lem:pattern} were checked state by state for all even $L\le40$ and several $n$ for each; the work of every access with $x\ge L^2/4+L/2+1$ was exactly $\frac52L-1$.

\end{document}